\documentclass[preprint]{rsl}

\title{Long-term meter wavelength variability study of Blazar J1415+1320 using the Ooty Radio Telescope}
\author{Sravani Vaddi, P. K. Manoharan, D. Anish Roshi}

\begin{document}

\maketitle

%
%

\begin{abstract}
 J1415+1320 is a well-studied blazar that exhibits strong flux density variability at a wide range of radio frequencies (2.4 $-$ 230~GHz).  In this letter, we present a variability study of this source at 327~MHz using the data obtained with the Ooty Radio Telescope taken during the period 1989 to 2018.  Two significant flares are detected at epochs 2007.6 and 2008.6.  These flares are also seen in the publicly available 15 and 37~GHz light curves but with a lead time of a few months. The fractional changes in the flux densities are larger at frequencies $>$ 15~GHz compared to those at 327~MHz and the spectral indices of the increased flux densities are flatter compared to the quiescent spectrum during these flares. These observed features are consistent with a model of uniformly expanding cloud of relativistic electrons or shock-in-jet model.  Our 327~MHz dataset also overlaps with a rare form of variability $-$ symmetric achromatic variability (SAV) $-$ seen at higher frequencies ($>$ 15~GHz) toward the source. SAV is possibly due to gravitational milli-lensing of the core emission by an intervening massive object and is expected to be detected at all frequencies \cite{Vedantham2017}. No variability in association with the SAV events is seen in the 327~MHz dataset; however, if SAV is due to the lensing of core emission alone then the expected variability is less than 3$\sigma$ uncertainty in our measurements.
\end{abstract}

\section{Introduction}
J1415+1320, a BL Lac has alluded to the astronomers for a very long time owing to its controversial properties.  Some of these properties include (a) apparent yet rare association of the active galactic nuclei (AGN) with an optical spiral host~\cite{McHardy1994}, (b) detection of a counter jet in a blazar type AGN~\cite{Perlman1994}, (c) its association to the class of compact symmetric objects as well as blazar \cite{Perlman1996}. Some of these controversies were addressed in \cite{Readhead2021} and conclude that J1415+1320 is actually a background object in the redshift range 0.247 $<$ z $<$ 0.5 and is not associated with the previously known spiral host.\\
The radio VLBI structure of J1415+1320 shows a two-sided, bent core-jet structure of size $\sim$ 110 mas ($<$ 1 kpc).  The core has an inverted spectrum ($\alpha$ $>$ 1) at frequencies between 1.4 $-$ 15 GHz \cite{Perlman1996}.  At frequencies above 15 GHz where the core emission dominates \cite{Readhead2021}, the core has a flat spectrum ($\alpha$ $\sim$ 0.001) (estimated using 15 and 37 GHz flux densities). The spectrum of the jet and the counter-jet are generally steep ($\alpha$ $<$ $-$1) but varies from being flat at the knots to steep in the more diffused region \cite{Perlman1996}. The total flux density of the source at 1.4 GHz is $\sim$ 1 Jy, dominated by emission from the jets, and increases to 8 Jy at 80 MHz \cite{Lister2018}. \\
The source exhibits strong variability in its radio light curve \cite{Vedantham2017}.  Variability in an AGN reveals crucial information on the size, structure, and dynamics of the radiating source down to scales that otherwise need extremely long baseline interferometers.  While variability in a source is a complicated feature that is not yet fully understood, several possible mechanisms were discussed in the literature broadly categorized into intrinsic and extrinsic phenomena. Intrinsic variability occurs due to, but not limited to, $-$ (a) shocks waves forming and propagating relativistically along the jets \cite{Marscher1985}, (b) magnetohydrodynamic instabilities in the jet \cite{Marscher2014}, (c)  variation in relativistic beaming as a result of viewing angle change in a twisted/bent jet \cite{Raiteri2017} (d) magnetic reconnection in turbulent jets \cite{Werner2016}. Variability is also observed owing to extrinsic phenomena such as refractive interstellar scintillation caused by large-scale irregularities in the interstellar medium (ISM) \cite{Quirrenbach1992}.\\
An interesting aspect of the radio light curve of J1415+1320 is that it displays a hitherto unrecognized form of variability, referred to as symmetric achromatic variability (SAV) \cite{Vedantham2017}. This variability is seen as a U-dip feature in the light curve, and is time-symmetric and achromatic over 15 to 234~GHz. This rare variability is possibly due to gravitational millilensing of the compact core emission in J1415+1320 by an intervening $10^2-10^5~M_{\odot}$ mass condensates \cite{Vedantham2017}. \\
So far most of the variability studies of J1415+1320 were made at frequencies $\ge$ 2.4~GHz. In this letter, we present long-term (1989.8 to 2017), low frequency (327~MHz) flux density observations of J1415+1320 taken using Ooty Radio Telescope (ORT), operated by the Radio Astronomy Centre, Tata Institute of Fundamental Research, India \cite{Swarup1971}. The details of observations and data reduction are given in Section~\ref{sec:data}. We report radio flux density variation of J1415+1320 at 327~MHz at epochs 2007.6 and 2008.6. Further, we use the data to investigate the achromatic nature of two previously reported SAV events down to 327~MHz, which is discussed in Section~\ref{sec:results}. The origin of the reported flux density variation is discussed in Section~\ref{sec:discussion} and our main conclusions are given in Section~\ref{sec:conclusions}.

\begin{figure}
    \centering
    \includegraphics[width=0.5\textwidth]{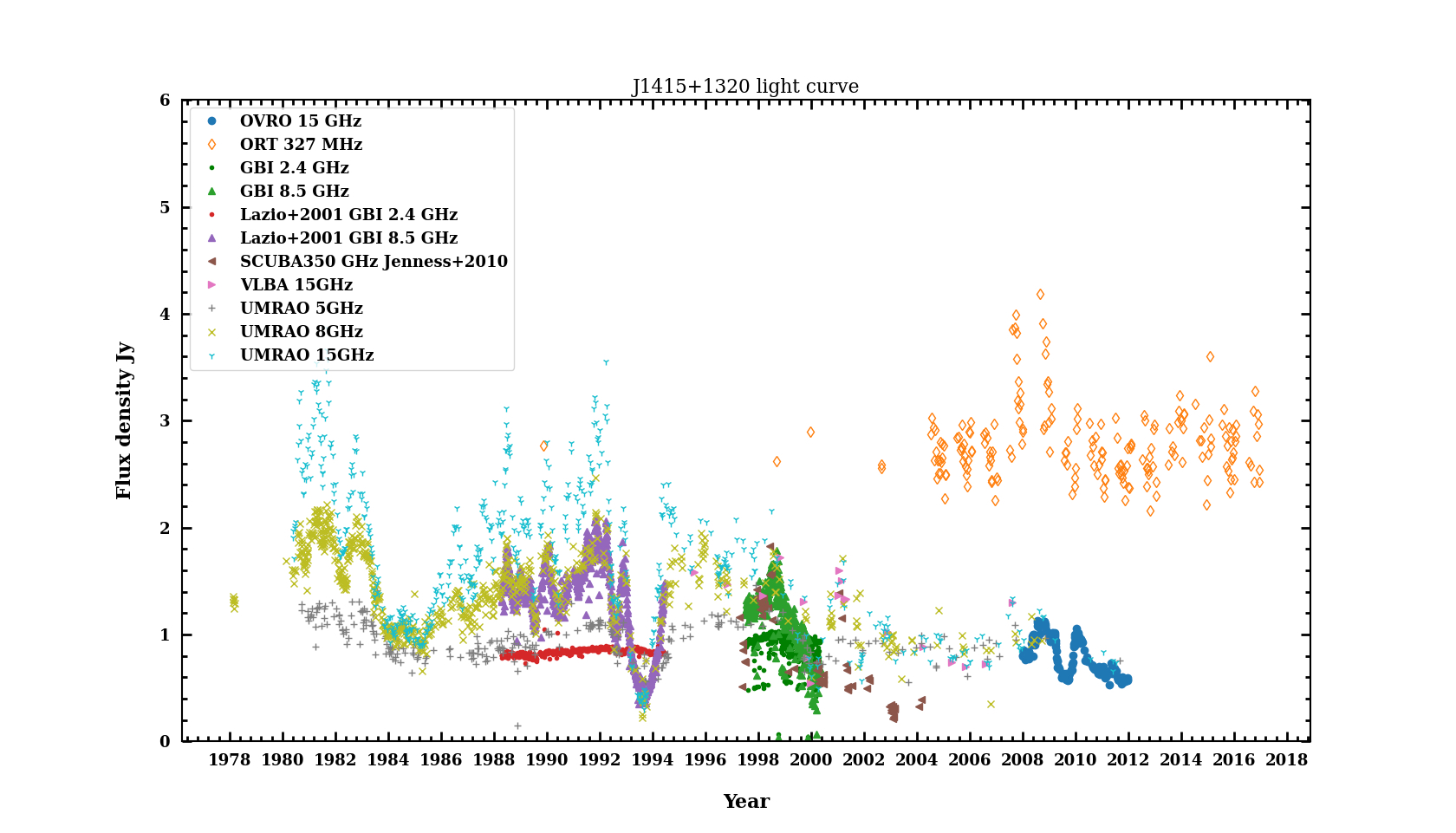}
    \caption{Radio light curve of J1415+1320 at different frequencies taken from the literature together with the ORT data. The data was compiled from the following $-$ Owens Valley Radio Observatory (OVRO) \cite{Richards2014}, Green Bank interferometer (GBI) \cite{Lazio2001}, submillimetre common-user bolometer array (SCUBA) \cite{Jenness2010}, very long baseleine array (VLBA) \cite{Lister2009}, University of Michigan Radio Astronomy Observatory (UMRAO) \cite{Aller2003}.}
    \label{fig:LC_allfreq}
\end{figure}

\section{Data}
\label{sec:data}
The data presented here comes from the extensive interplanetary scintillation (IPS) observations made with the ORT during the period 1989$-$2018. About 100 compact radio sources of angular size less than 250~mas were observed per day at 327~MHz as part of the IPS campaign. The main purpose of these IPS observations was to investigate the three-dimensional distributions of the solar wind density, turbulence, and speed at short time intervals of a few days as well as at different phases of solar activity \cite{Manoharan2012,Manoharan2017}. The IPS measurements on each source were made within the solar elongation range of about $\varepsilon \thickapprox \pm$ 60 degree with respect to the Sun, which spanned over an observing period of about 4 months. Typically a source was observed for about 2–3 minutes and the same source was likely observed more than once in a day at different hour-angles. Each observing session also included observations of several flux density calibrators distributed almost uniformly between the start and end of the session. \\
In this paper, the above discussed large IPS database was employed to study the variation of the total flux density of J1415+1320 over many years. The data for J1415+1320 and the control source B1345+125 are made available online in VizieR.

\section{Results}
\label{sec:results}
\begin{figure}
    \centering
    \includegraphics[width=\columnwidth]{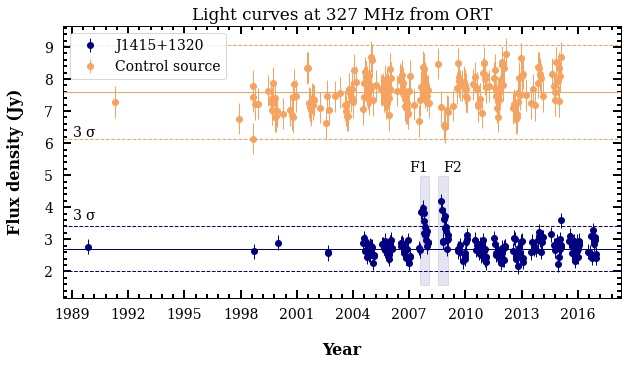}
    \caption{327~MHz light curve of J1415+1320 (blue) and control source B1345+125 (orange) obtained from ORT IPS measurements.  The solid line and dashed line mark the mean and 3$\sigma$ standard deviation. F1 and F2 mark the flares at epoch 2007.6 and 2008.6 respectively.}
    \label{fig:327MHz_LC}
\end{figure}

\begin{figure}[!h]
    \centering
     \includegraphics[width=\columnwidth]{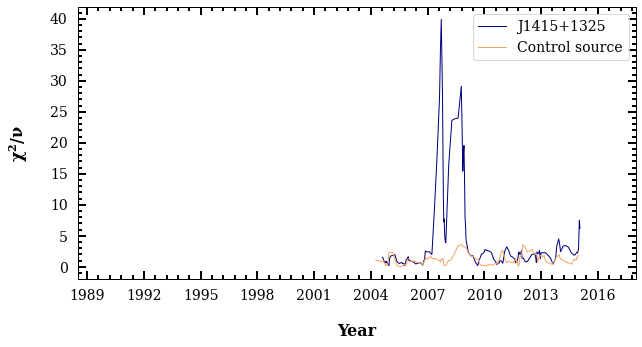}
    \caption{Chi-square test statistic normalized by the degrees of freedom ($\nu$) plotted for different epochs (see text).}
    \label{fig:chisq}
\end{figure}

Figure \ref{fig:LC_allfreq} shows the radio light curves of J1415+1320 spanning nearly 40 years (1978$-$2018). The data at several frequencies ranging from 2.4 to 350~GHz was compiled from various monitoring programs found in the literature \cite{Lazio2001,Aller2003,Lister2009,Jenness2010,Richards2014}.  The source exhibits variability at all frequencies.  Strong variability is seen at higher frequencies compared to lower frequencies.  The variability features seen at 15~GHz are followed at 8 and 5~GHz albeit at a lower level.  Further, for the 15~GHz waveband, the strength of variability generally decreases over time.\\
We have included the 327~MHz radio light curve over the period 1989.8 to 2017 in Figure \ref{fig:LC_allfreq} for comparison. The data used for the 327~MHz light curve contains 1953 individual observations, which are obtained after excluding observations taken at small solar elongations (i.e., $\varepsilon \leq$ 5 degree) to avoid confusion caused by the telescope side lobe pointing at the Sun. Each data point represents the average of nearly 10 to 15 consecutive observations taken around an epoch (a few days). For years 1989, 1998, 1999, and 2002, the available number of observations is limited to only about 10 and their average is plotted. \\
In Figure \ref{fig:327MHz_LC}, we reproduce the 327~MHz light curve of J1415+1320 along with the flux density measurements of the control source B1345+125. The control source B1345+125 is a compact ($<$ 100 mas) steep spectrum source \cite{Lister2003}, monitored during a similar time period of observation of J1415+1320. The angular distance of the control source from J1415+1320 is about 10 deg. Both sources display strong IPS (scintillation index of the order of unity \cite{Manoharan1995, Manoharan2012}) and so the long-term epoch-to-epoch variations in both sources are dominated by scintillation. In other words, the normalized $\chi^2$ variability test as defined in \cite{Kesteven1976} gives a value close to unity when applied to the two datasets. The flux density of J1415+1320 interestingly shows a significant increase at epochs 2007.6 and 2008.6 (see Fig \ref{fig:327MHz_LC}); we refer to these flares as F1 and F2 respectively. The mean flux density of J1415+1320, estimated after excluding the data from the two epochs 2007.6 and 2008.6, is 2.70~Jy and the standard deviation $\sigma$, of flux density variation is 0.23~Jy.  During flare F1, the flux density increased rapidly from 2.66~Jy to 3.99~Jy, and during F2, the flux density increased from  2.9~Jy to 4.19~Jy; an increase of $\sim$ 45\%.  No variation in flux density is observed on the control source B1345+125; the mean and standard deviation for the control source are 7.6~Jy and 0.5~Jy respectively.  The absence of flux density variation in the control source around epochs 2007.6 and 2008.6 rules out any instrumental effects or IPS causing the variation in the flux density of J1415+1320. \\
The maximum flux densities during flares F1 and F2 are 3.85~Jy (17$\sigma$) and 4.19~Jy (18$\sigma$) respectively. We perform $\chi^2$ test on the target and the control source to further quantify the significance of the flare. Figure \ref{fig:chisq} shows the chi-square test statistic per degree of freedom as a function of epoch.  We define chi-square statistic $ \chi^2 = \sum_{i=1}^{N}\frac{\left(S_i-\bar{S}\right)^2}{\sigma_i^2}$ where S$_i$ and $\sigma_i$ are the flux density and corresponding measurement error at $i^{th}$ epoch and $\bar{S}$ is the mean flux density. The magnitude of $\sigma_i$ is determined by the IPS and is equal to the standard deviation (0.23~Jy) of the flux density variation away from the flares. $\bar{S} = 2.7$~Jy is the mean flux density estimated above. In Figure \ref{fig:chisq}, $ \chi^2$ is computed for the data points in a moving window with $N=4$ values and plotted against the epoch. The chi-square value is $>10$ during the flares and the p-value is less than 0.003. Thus we conclude that the flares detected at 2007.6 and 2008.6 in the J1415+1320 radio light curve at 327~MHz are statistically significant. \\
Figure \ref{fig:327MHz_flare_region} shows a close-up of the radio light curve from epoch 2004 to 2017. The flares F1 and F2 are marked in the figure. During the first flare F1, the flux density increases rapidly from 2.66~Jy to 3.99~ Jy (a variation of 1.33$\pm$0.22~Jy) in four days (all times given in this paper is relative to the observer).  The flux density falls gradually in the next $\sim$ 84 days and reaches a low of 2.78~Jy on 2007.97.  The second flare F2 is observed on 2008.67 where the flux density increases to 4.19~Jy. Unfortunately, no observations are present after the decay of the first flare and the onset of the second flare and so we are unable to estimate the rise time of this flare. The flux density of F2 falls in about 134 days and returns to 2.71~Jy on 2009.03.  During F2, there is an abrupt fall in the flux density from 3.91~Jy to 2.92~Jy ($>$ 3$\sigma$ variation) on 2008.82.  This lasted for about 11 days and then increased to 3.63~Jy on 2008.85.  


\section{Discussion}
\label{sec:discussion}

\subsection{Flares F1, F2}
\begin{figure}
    \centering
    \includegraphics[width=\columnwidth]{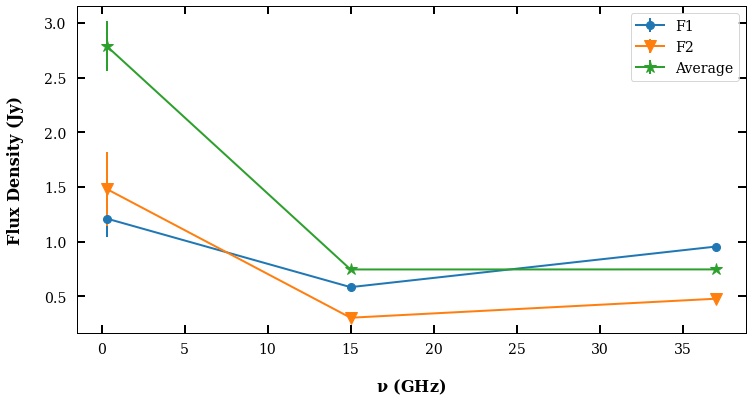}
    \caption{Spectrum of the increased flux density during the flares F1 and F2 i.e. difference between maximum (representing the peak of the flare) and minimum (representing the quiescent state) flux density, and the average flux density.}
    \label{fig:spec}
\end{figure}

  We compare our data with available observations at higher frequencies. We find light curves at 15~GHz OVRO data and 37~GHz Mets\"{a}hovi radio observatory monitoring program overlapping with our low frequency observations.  Figure \ref{fig:327MHz_flare_region} shows light curves measured at these high frequencies.  An event with a rapid increase followed by a gradual decay nearer to the epochs of F1 and F2 is seen in the higher frequency light curves. The peak of the flare F1 at 327~MHz occurred $\sim 115$~days after the epoch at which the flux density peaked at 37~GHz. The time delay between 327~MHz and 15~GHz flares is $\sim 49$~days. For F2, the time delays are $\sim 15$ and $\sim 73$~days relative to 15 and 37~GHz flux density maxima. \\
 The fractional increase in flux densities at higher frequencies is $\sim$ 80\% for both the flares; more than 1.6 compared to flux density increase obtained from the 327~MHz data. Also, the decay time scale at higher frequencies is larger by a factor of $\sim$ 2 compared to the decay time scale observed at 327~MHz. Further the spectrum of the increased flux density between 327 and 15~GHz is flatter ($\alpha=-0.2$), compared to the quiescent flux density spectrum ($\alpha=-0.3$; see Fig.~\ref{fig:spec}). For frequencies above 15~GHz, the increased flux density and quiescent values have similar spectral indices (see Fig.~\ref{fig:spec}). All these indicate that the flares might have originated outside the optically thick region of the source;  the steeping of the flux density at 327~MHz during the quiescent state is due to increased contribution from the jet (and/or counter-jet) in J1415+1320 \cite{Perlman1996}. The observed flux density variation and the time evolution of the flare are generally consistent with a uniformly expanding cloud of relativistic electrons or shock-in-jet model \cite{van der Laan1966, Marscher1985}.  
 
 \subsection{Symmetric Achromatic Variability}

\begin{figure}
    \centering
    \includegraphics[width=\columnwidth]{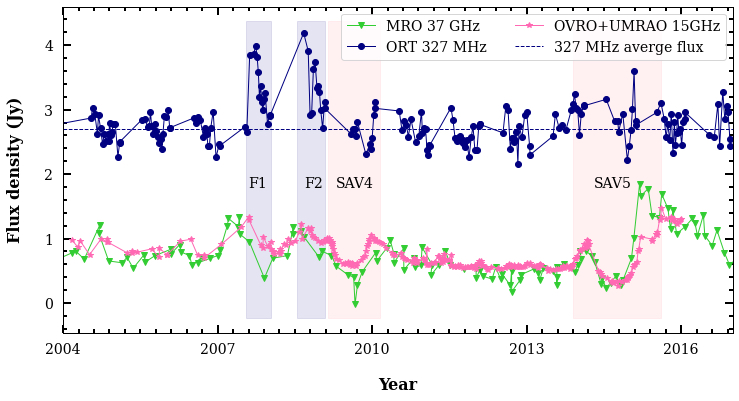}
    \caption{327~MHz, 15 and 37~GHz light curves (from OVRO, UMRAO, and MRO) of J1415+1320 covering period between 2004$-$2017. The two flares are indicated by shaded regions F1 and F2 and the two SAVs found in 15~GHz light curve and reported in \cite{Peirson2022} are marked as shaded SAV4 and SAV5 following their notation.}
    \label{fig:327MHz_flare_region}
\end{figure}

The flux density measurements at 327~MHz overlap with two symmetric achromatic variability (SAV) identified at higher frequencies ($>$ 15~GHz) by \cite{Vedantham2017}. These SAV events are marked in Figure \ref{fig:327MHz_flare_region} as SAV4 and SAV5 following the notation from \cite{Peirson2022}.  The SAVs can be seen as a U$-$dip feature in the 15 and 37~GHz light curves. No variability of flux density within $\pm$0.7~Jy (3$\sigma$) is seen at 327~MHz during these two SAV events.  \\
The SAV is possibly due to gravitational milli-lensing of the compact core emission by intervening mass condensates \cite{Vedantham2017, Peirson2022}. Therefore the variability is expected to be achromatic and would have been seen in the 327~MHz data.  The spectral index of the core component is inverted with $\alpha=+1.7$ \cite{Perlman1996}. Thus the expected core flux density at 327~MHz is 1.2 mJy and the fractional change of $\sim$ 80\% inferred from higher frequency light curves during the SAV is much smaller than the measurement uncertainty of flux density at 327~MHz. Thus the non-detection of any variability at 327~MHz is consistent with the gravitational lensing model proposed to explain SAV. 

\section{Conclusion}
\label{sec:conclusions}
We studied the 327~MHz multi-epoch data of J1415+1320, a blazar known for its high variability at higher frequencies.  The total flux density time series was obtained from the IPS database from the ORT and covers the period between 1989 to 2018, although much of the data points exist after 2004. We report significant variability at two epochs 2007.6 (flare F1) and 2008.6 (flare F2) and establish the significance of these flares through statistical analysis. During these flares, the flux density rises rapidly, reaches a maximum, and then decreases gradually. Both F1 and F2 are seen in the overlapping datasets at 15 and 37~GHz, but with a time delay of a few months.  The spectral index of the increased flux density during the flare is flatter than the quiescent spectral index, indicating the flare is likely to be associated with activity in a region with smaller optical depth. Our dataset also overlaps with two SAV events (SAV4 and SAV5) identified by \cite{Vedantham2017, Peirson2022}. No variability of flux density is seen within $\pm$ 0.7~Jy (3$\sigma$) level at 327~MHz during these events. Our non-detection of variability is consistent with a gravitational lensing origin for the SAVs if only the core flux density is affected by the lensing phenomenon.   

\section{Acknowledgements}
Thanks to the anonymous referee for insightful suggestions which helped improve the manuscript. We acknowledge the help from the Radio Astronomy Centre, Ooty staff for their help during the observations.  This research has made use of data from the University of Michigan Radio Astronomy Observatory which has been supported by the University of Michigan and by a series of grants from the National Science Foundation, most recently AST-0607523.

%
%
\noindent\small
The authors are affiliated with the following institutions:\\
Arecibo Observatory, Arecibo, Puerto Rico 00612, USA\\
University of Central Florida, 4000 Central Florida Blvd, Orlando, Florida 32816 USA\\
Sravani Vaddi; e-mail: sravani.vaddi@gmail.com

\begin{thebibliography}{}
\makeatletter
\relax
\def\mn@urlcharsother{\let\do\@makeother \do\$\do\&\do\#\do\^\do\_\do\%\do\~}
\def\mn@doi{\begingroup\mn@urlcharsother \@ifnextchar [ {\mn@doi@}
  {\mn@doi@[]}}
\def\mn@doi@[#1]#2{\def\@tempa{#1}\ifx\@tempa\@empty \href
  {http://dx.doi.org/#2} {doi:#2}\else \href {http://dx.doi.org/#2} {#1}\fi
  \endgroup}
\def\mn@eprint#1#2{\mn@eprint@#1:#2::\@nil}
\def\mn@eprint@arXiv#1{\href {http://arxiv.org/abs/#1} {{\tt arXiv:#1}}}
\def\mn@eprint@dblp#1{\href {http://dblp.uni-trier.de/rec/bibtex/#1.xml}
  {dblp:#1}}
\def\mn@eprint@#1:#2:#3:#4\@nil{\def\@tempa {#1}\def\@tempb {#2}\def\@tempc
  {#3}\ifx \@tempc \@empty \let \@tempc \@tempb \let \@tempb \@tempa \fi \ifx
  \@tempb \@empty \def\@tempb {arXiv}\fi \@ifundefined
  {mn@eprint@\@tempb}{\@tempb:\@tempc}{\expandafter \expandafter \csname
  mn@eprint@\@tempb\endcsname \expandafter{\@tempc}}}

\bibitem[\protect\citeauthoryear{{Aller}, {Aller}  \& {Hughes}}{{Aller}
  et~al.}{1985}]{Aller1985}
{Aller} H.~D.,  {Aller} M.~F.,   {Hughes} P.~A.,  1985, \mn@doi [\apj]
  {10.1086/163610}, \href
  {https://ui.adsabs.harvard.edu/abs/1985ApJ...298..296A} {298, 296}

\bibitem[\protect\citeauthoryear{{Aller}, {Aller}  \& {Hughes}}{{Aller}
  et~al.}{2003}]{Aller2003}
{Aller} M.~F.,  {Aller} H.~D.,   {Hughes} P.~A.,  2003, in {Zensus} J.~A.,
  {Cohen} M.~H.,   {Ros} E.,  eds,  Astronomical Society of the Pacific
  Conference Series Vol. 300, Radio Astronomy at the Fringe. p.~159

\bibitem[\protect\citeauthoryear{{Jenness}, {Robson}  \& {Stevens}}{{Jenness}
  et~al.}{2010}]{Jenness2010}
{Jenness} T.,  {Robson} E.~I.,   {Stevens} J.~A.,  2010, \mn@doi [\mnras]
  {10.1111/j.1365-2966.2009.15716.x}, \href
  {https://ui.adsabs.harvard.edu/abs/2010MNRAS.401.1240J} {401, 1240}

\bibitem[\protect\citeauthoryear{{Kesteven}, {Bridle}  \& {Brandie}}{{Kesteven}
  et~al.}{1976}]{Kesteven1976}
{Kesteven} M.~J.~L.,  {Bridle} A.~H.,   {Brandie} G.~W.,  1976, \mn@doi [\aj]
  {10.1086/111971}, \href
  {https://ui.adsabs.harvard.edu/abs/1976AJ.....81..919K} {81, 919}

\bibitem[\protect\citeauthoryear{{Lazio}, {Waltman}, {Ghigo}, {Fiedler},
  {Foster}  \& {Johnston}}{{Lazio} et~al.}{2001}]{Lazio2001}
{Lazio} T. J.~W.,  {Waltman} E.~B.,  {Ghigo} F.~D.,  {Fiedler} R.~L.,  {Foster}
  R.~S.,   {Johnston} K.~J.,  2001, \mn@doi [\apjs] {10.1086/322531}, \href
  {https://ui.adsabs.harvard.edu/abs/2001ApJS..136..265L} {136, 265}

\bibitem[\protect\citeauthoryear{{Lister}, {Kellermann}, {Vermeulen}, {Cohen},
  {Zensus}  \& {Ros}}{{Lister} et~al.}{2003}]{Lister2003}
{Lister} M.~L.,  {Kellermann} K.~I.,  {Vermeulen} R.~C.,  {Cohen} M.~H.,
  {Zensus} J.~A.,   {Ros} E.,  2003, \mn@doi [\apj] {10.1086/345666}, \href
  {https://ui.adsabs.harvard.edu/abs/2003ApJ...584..135L} {584, 135}

\bibitem[\protect\citeauthoryear{{Lister} et~al.,}{{Lister}
  et~al.}{2009}]{Lister2009}
{Lister} M.~L.,  et~al., 2009, \mn@doi [\aj] {10.1088/0004-6256/137/3/3718},
  \href {https://ui.adsabs.harvard.edu/abs/2009AJ....137.3718L} {137, 3718}

\bibitem[\protect\citeauthoryear{{Manoharan}}{{Manoharan}}{2012}]{Manoharan2012}
{Manoharan} P.~K.,  2012, \mn@doi [\apj] {10.1088/0004-637X/751/2/128}, \href
  {https://ui.adsabs.harvard.edu/abs/2012ApJ...751..128M} {751, 128}

\bibitem[\protect\citeauthoryear{{Manoharan}, {Ananthakrishnan}, {Dryer},
  {Detman}, {Leinbach}, {Kojima}, {Watanabe}  \& {Kahn}}{{Manoharan}
  et~al.}{1995}]{Manoharan1995}
{Manoharan} P.~K.,  {Ananthakrishnan} S.,  {Dryer} M.,  {Detman} T.~R.,
  {Leinbach} H.,  {Kojima} M.,  {Watanabe} T.,   {Kahn} J.,  1995, \mn@doi
  [\solphys] {10.1007/BF00670233}, \href
  {https://ui.adsabs.harvard.edu/abs/1995SoPh..156..377M} {156, 377}

\bibitem[\protect\citeauthoryear{{Manoharan}, {Subrahmanya}  \&
  {Chengalur}}{{Manoharan} et~al.}{2017}]{Manoharan2017}
{Manoharan} P.~K.,  {Subrahmanya} C.~R.,   {Chengalur} J.~N.,  2017, \mn@doi
  [Journal of Astrophysics and Astronomy] {10.1007/s12036-017-9435-z}, \href
  {https://ui.adsabs.harvard.edu/abs/2017JApA...38...16M} {38, 16}

\bibitem[\protect\citeauthoryear{{Marscher}}{{Marscher}}{2014}]{Marscher2014}
{Marscher} A.~P.,  2014, \mn@doi [\apj] {10.1088/0004-637X/780/1/87}, \href
  {https://ui.adsabs.harvard.edu/abs/2014ApJ...780...87M} {780, 87}

\bibitem[\protect\citeauthoryear{{Marscher} \& {Gear}}{{Marscher} \&
  {Gear}}{1985}]{Marscher1985}
{Marscher} A.~P.,  {Gear} W.~K.,  1985, \mn@doi [\apj] {10.1086/163592}, \href
  {https://ui.adsabs.harvard.edu/abs/1985ApJ...298..114M} {298, 114}

\bibitem[\protect\citeauthoryear{{McHardy}, {Merrifield}, {Abraham}  \&
  {Crawford}}{{McHardy} et~al.}{1994}]{McHardy1994}
{McHardy} I.~M.,  {Merrifield} M.~R.,  {Abraham} R.~G.,   {Crawford} C.~S.,
  1994, \mn@doi [\mnras] {10.1093/mnras/268.3.681}, \href
  {https://ui.adsabs.harvard.edu/abs/1994MNRAS.268..681M} {268, 681}

\bibitem[\protect\citeauthoryear{{Peirson} et~al.,}{{Peirson}
  et~al.}{2022}]{Peirson2022}
{Peirson} A.~L.,  et~al., 2022, arXiv e-prints, \href
  {https://ui.adsabs.harvard.edu/abs/2022arXiv220101110P} {p. arXiv:2201.01110}

\bibitem[\protect\citeauthoryear{{Perlman}, {Stocke}, {Shaffer}, {Carilli}  \&
  {Ma}}{{Perlman} et~al.}{1994}]{Perlman1994}
{Perlman} E.~S.,  {Stocke} J.~T.,  {Shaffer} D.~B.,  {Carilli} C.~L.,   {Ma}
  C.,  1994, \mn@doi [\apjl] {10.1086/187277}, \href
  {https://ui.adsabs.harvard.edu/abs/1994ApJ...424L..69P} {424, L69}

\bibitem[\protect\citeauthoryear{{Perlman}, {Carilli}, {Stocke}  \&
  {Conway}}{{Perlman} et~al.}{1996}]{Perlman1996}
{Perlman} E.~S.,  {Carilli} C.~L.,  {Stocke} J.~T.,   {Conway} J.,  1996,
  \mn@doi [\aj] {10.1086/117922}, \href
  {https://ui.adsabs.harvard.edu/abs/1996AJ....111.1839P} {111, 1839}

\bibitem[\protect\citeauthoryear{{Qian}, {Quirrenbach}, {Witzel}, {Krichbaum},
  {Hummel}  \& {Zensus}}{{Qian} et~al.}{1991}]{Qian1991}
{Qian} S.~J.,  {Quirrenbach} A.,  {Witzel} A.,  {Krichbaum} T.~P.,  {Hummel}
  C.~A.,   {Zensus} J.~A.,  1991, \aap, \href
  {https://ui.adsabs.harvard.edu/abs/1991A&A...241...15Q} {241, 15}

\bibitem[\protect\citeauthoryear{{Quirrenbach} et~al.,}{{Quirrenbach}
  et~al.}{1992}]{Quirrenbach1992}
{Quirrenbach} A.,  et~al., 1992, \aap, \href
  {https://ui.adsabs.harvard.edu/abs/1992A&A...258..279Q} {258, 279}

\bibitem[\protect\citeauthoryear{{Raiteri} et~al.,}{{Raiteri}
  et~al.}{2017}]{Raiteri2017}
{Raiteri} C.~M.,  et~al., 2017, \mn@doi [\nat] {10.1038/nature24623}, \href
  {https://ui.adsabs.harvard.edu/abs/2017Natur.552..374R} {552, 374}

\bibitem[\protect\citeauthoryear{{Readhead} et~al.,}{{Readhead}
  et~al.}{2021}]{Readhead2021}
{Readhead} A.~C.~S.,  et~al., 2021, \mn@doi [\apj] {10.3847/1538-4357/abd08c},
  \href {https://ui.adsabs.harvard.edu/abs/2021ApJ...907...61R} {907, 61}

\bibitem[\protect\citeauthoryear{{Richards}, {Hovatta}, {Max-Moerbeck},
  {Pavlidou}, {Pearson}  \& {Readhead}}{{Richards} et~al.}{2014}]{Richards2014}
{Richards} J.~L.,  {Hovatta} T.,  {Max-Moerbeck} W.,  {Pavlidou} V.,  {Pearson}
  T.~J.,   {Readhead} A.~C.~S.,  2014, \mn@doi [\mnras]
  {10.1093/mnras/stt2412}, \href
  {https://ui.adsabs.harvard.edu/abs/2014MNRAS.438.3058R} {438, 3058}

\bibitem[\protect\citeauthoryear{{Sironi}, {Petropoulou}  \&
  {Giannios}}{{Sironi} et~al.}{2015}]{Sironi2015}
{Sironi} L.,  {Petropoulou} M.,   {Giannios} D.,  2015, \mn@doi [\mnras]
  {10.1093/mnras/stv641}, \href
  {https://ui.adsabs.harvard.edu/abs/2015MNRAS.450..183S} {450, 183}

\bibitem[\protect\citeauthoryear{{Swarup} et~al.,}{{Swarup}
  et~al.}{1971}]{Swarup1971}
{Swarup} G.,  et~al., 1971, \mn@doi [Nature Physical Science]
  {10.1038/physci230185a0}, \href
  {https://ui.adsabs.harvard.edu/abs/1971NPhS..230..185S} {230, 185}

\bibitem[\protect\citeauthoryear{{Vedantham} et~al.,}{{Vedantham}
  et~al.}{2017}]{Vedantham2017}
{Vedantham} H.~K.,  et~al., 2017, \mn@doi [\apj] {10.3847/1538-4357/aa7741},
  \href {https://ui.adsabs.harvard.edu/abs/2017ApJ...845...90V} {845, 90}

\bibitem[\protect\citeauthoryear{{Werner}, {Uzdensky}, {Cerutti}, {Nalewajko}
  \& {Begelman}}{{Werner} et~al.}{2016}]{Werner2016}
{Werner} G.~R.,  {Uzdensky} D.~A.,  {Cerutti} B.,  {Nalewajko} K.,   {Begelman}
  M.~C.,  2016, \mn@doi [\apjl] {10.3847/2041-8205/816/1/L8}, \href
  {https://ui.adsabs.harvard.edu/abs/2016ApJ...816L...8W} {816, L8}

\bibitem[\protect\citeauthoryear{{Wilkinson}, {Polatidis}, {Readhead}, {Xu}  \&
  {Pearson}}{{Wilkinson} et~al.}{1994}]{Wilkinson1994}
{Wilkinson} P.~N.,  {Polatidis} A.~G.,  {Readhead} A.~C.~S.,  {Xu} W.,
  {Pearson} T.~J.,  1994, \mn@doi [\apjl] {10.1086/187518}, \href
  {https://ui.adsabs.harvard.edu/abs/1994ApJ...432L..87W} {432, L87}

\bibitem[\protect\citeauthoryear{{van der Laan}}{{van der
  Laan}}{1966}]{vanLaan1966}
{van der Laan} H.,  1966, \mn@doi [\nat] {10.1038/2111131a0}, \href
  {https://ui.adsabs.harvard.edu/abs/1966Natur.211.1131V} {211, 1131}

\makeatother
\end{thebibliography}
\end{document}